\documentclass[sigconf,edbt]{acmart-edbt2021}

\def\BibTeX{{\rm B\kern-.05em{\sc i\kern-.025em b}\kern-.08em
    T\kern-.1667em\lower.7ex\hbox{E}\kern-.125emX}}

\usepackage{booktabs} 
\usepackage{color} 
\usepackage{subcaption}
\usepackage{array} 
\usepackage[normalem]{ulem}

\setcopyright{rightsretained}

\acmDOI{}

\acmISBN{978-3-89318-084-4}

\acmConference[EDBT 2021]{24th International Conference on Extending Database Technology (EDBT)}{March 23-26, 2021}{Nicosia, Cyprus} 
\acmYear{2021}

\begin{document}
\title{Simplifying Impact Prediction for Scientific Articles}

\author{Thanasis Vergoulis}
\affiliation{%
  \institution{IMSI, ATHENA RC}
}
\email{vergoulis@athenarc.gr}

\author{Ilias Kanellos}
\affiliation{%
  \institution{IMSI, ATHENA RC}
}
\email{ilias.kanellos@athenarc.gr}

\author{Giorgos Giannopoulos}
\affiliation{%
  \institution{IMSI, ATHENA RC}
}
\email{giann@athenarc.gr}

\author{Theodore Dalamagas}
\affiliation{%
  \institution{IMSI, ATHENA RC}
}
\email{dalamag@athenarc.gr}

\renewcommand{\shortauthors}{}

\begin{abstract}
Estimating the expected impact of an article is valuable for various applications (e.g., article/cooperator recommendation). Most existing approaches attempt to predict the exact number of citations each article will receive in the near future, however this is a difficult regression analysis problem. Moreover, most approaches rely on the existence of rich metadata for each article, a requirement that cannot be adequately fulfilled for a large number of them. In this work, we take advantage of the fact that solving a simpler machine learning problem, that of classifying articles based on their expected impact, is adequate for many real world applications and we propose a simplified model that can be trained using minimal article metadata. Finally, we examine various configurations of this model and evaluate their effectiveness in solving the aforementioned classification problem.  
\end{abstract}

%
%


\maketitle

\section{Introduction}
\label{sec:intro}

Predicting the attention a scientific article will attract in the next few years by other articles, i.e., estimating its expected \emph{impact}\footnote{Since scientific impact has several aspects \cite{bollen2009principal}, the term can be defined in diverse ways. In this work, we focus on the definition provided in Section~\ref{sec:approach-prel}.}, is very useful for many applications. For example, consider a recommendation system, which suggests articles to researchers based on their interests. Due to the large growth rate in the number of published research works~\cite{growth2}, a large number of articles will be retrieved for almost any subject of interest. However, not all of them will be of equal importance. The recommendation system could leverage the
expected impact of papers to suggest only the most important
works to the user and avoid overwhelming her with a large number of
trivial options. The benefits would be similar for other relevant applications, such as expert finding, collaboration recommendation, etc.

Several approaches, which attempt to predict the exact number of citations articles will receive in the next few years, have been proposed in the literature (see Section~\ref{sec:related} for indicative examples). However, this is an extremely difficult regression analysis problem, due to the many factors (some of which are hard to quantify) that may affect the impact of an article (details in Section~\ref{sec:approach-prob}). Fortunately, in practice, for many applications, knowing the exact number of future citations is not critical. For instance, in the case of the recommendation system, it is important that the system distinguish the `impactful' works from those that are of lesser importance; all impactful works will be interesting regardless of the exact number of citations they will receive. 

In addition, most existing approaches rely on rich article metadata (e.g., authors, venue, topics). Unfortunately, the available information for many articles in the relevant data sources (e.g., Crossref) is erroneous or incomplete, complicating the learning process of such approaches and creating risks for their effectiveness. Moreover, even when the required metadata are available, the generation of the corresponding machine learning features from them may be extremely time-consuming or even difficult to be implemented (details in Section~\ref{sec:approach-feat}). 

In this work, our objective is to take advantage of the previous observations in an attempt to guide and facilitate the work of researchers and developers working on applications that can benefit from predicting the expected impact of scientific articles. In particular, we propose a simplified machine learning approach which is based on the binary classification of articles in two categories (`impactful' / `impactless') according to their expected impact. In addition, we propose the use of a particular set of features that rely on minimal metadata for each article (only its publication year and its previous citations). We argue that this simpler approach is adequate, significantly easier
to implement, and can benefit many applications that
require the estimation of the expected impact of articles. Finally, we perform experiments to investigate the effectiveness of this approach using various well-established classifiers. In our experimental setup we
seriously take into consideration the fact that our problem 
is imbalanced by nature, both to carefully select the
appropriate evaluation measures and to examine some classification 
approaches that are particularly
tailored to such scenarios.

\section{Our approach}
\label{sec:approach}

\subsection{Preliminaries}
\label{sec:approach-prel}

Scientific articles always include a list of references to other works and the referenced articles describe work related to the referencing article (e.g., preliminaries, competitive approaches). As a result, the inclusion of an article in the reference list of another (i.e., the one \emph{citing} it) implies that the latter gives credit to the former\footnote{Note that the ``amount'' of credit may be significantly different for each referenced work and that, in some cases, it may also have a negative sign (when the referencing work criticizes the referenced one). }. Based on this view, counting the number of distinct articles that include an article of interest in their reference list (i.e., counting its \emph{citations}) is considered to be an indicator of its impact in the scientific community. Of course, there are also many other aspects of scientific impact~\cite{bollen2009principal}, however the focus of this work is on this type of citation-based expected impact. In particular, we focus on the \emph{expected impact} of an article at a given time point, which can be defined as follows: 

\begin{definition}[Expected Article Impact]
Consider an article $a$ and a time point $t$. Then, $i(a,t)$, the (expected) impact of $a$ at $t$, is calculated as the number of citations that $a$ will receive during the period $[t,t+y]$, where $y$ is a problem parameter, which defines a \emph{future period} of interest.  
\end{definition}

It should be noted that the problem parameter $y$ can be configured based on the characteristics of the dataset used. The optimal option typically depends on the
citation dynamics of the scientific fields covered by the dataset. However, $y=3$ or $y=5$ are two reasonable and very common configurations. Finally, it should be highlighted that the expected impact of an article can only be measured in retrospect, i.e., by monitoring the citations that the article receives $y$ years after the time point of reference.

\subsection{Problem definition}
\label{sec:approach-prob}

Considering the expected impact of articles can be useful for many applications. This is why there is a line of work of methods that attempt to predict the exact impact of each article, i.e., the exact number of citations it is going to receive in the following few years (see Section~\ref{sec:related}). However, this is a difficult regression analysis problem for many reasons. First of all, there are many factors that may affect the number of citations an article will receive in the future. These factors are related to the quality of the work, the hype of its topic, the prestige of its authors or its venue, the dissemination effort that will be made in social media, to name only a few. Also, to make matters worse, many of these factors cannot be easily quantified without losing important information (e.g., due to dimensionality reduction reasons in one-hot encodings), affecting the accuracy of the approaches.

Additionally, in practice, many of the aforementioned applications do not require the prediction of the exact number of future citations for each article. It is
sufficient for them to simply distinguish between `impactful' (to-be) and `impactless' articles. This type of problem is easier and, thus, a traditional classification approach is likely to achieve adequate effectiveness in solving it. Hence, 
in this work, we focus on a binary impact-based article 
classification problem that can be formulated as follows:

\begin{definition}[Impact-based article classification]
Consider a collection of scientific articles $A$ and a time point $t$ and let $\bar{i}=\sum_{a\in A} i(a,t)/|A|$. Then, the objective is to classify each $a \in A$ in one of two classes: in the class of `impactful' articles, if $i(a,t)> \bar{i}$ and to the class of `impactless' articles, otherwise.
\end{definition}

In other words, our objective is to identify articles that receive an above-average number of citations, to classify them as `impactful' and the rest as `impactless'. Note that this intuitive distinction is equivalent with the first iteration of the Head/Tail Breaks clustering algorithm, which is tailored for heavy tailed distributions, like the citation distribution of articles~\cite{barabasi2016network} (a small number of articles receive an extremely large number of citations). 

An important matter that should be highlighted is that this classification problem is \emph{imbalanced} by nature. Due to the fact that the citation distribution of articles is long-tailed, most articles have an impact (i.e., number of citations) well below average. Consequently, the class of `impactful' articles will always be a minority in the collection (the so-called `head' of the citation distribution). This is important for two reasons; first of all, it affects the correct choice of 
evaluation measures in the experimental setup. For example, using the accuracy (i.e., the ratio of true positives to the complete set) is problematic: a trivial classifier that would always assign all articles to the `impactless' class will always achieve a good performance according to this measure. For this reason, alternative measures like precision, recall, and F1 of the minority class (i.e., the class of `impactful' articles) should be used instead. Unfortunately, part of the previous literature (e.g., \cite{su2020prediction}) overlooks this issue making it difficult to evaluate the real effectiveness of the corresponding proposed approaches.  

\subsection{The proposed feature selection}
\label{sec:approach-feat}

Many existing machine learning approaches rely on the existence of various article metadata such as its publication year, author list, venue, main topics, citations etc. Although nowadays a large portion of such data becomes available through open scholarly graphs \cite{orkg,oa-graph} or datasets (e.g., DBLP, Crossref), there are many articles for which important information is erroneous, incomplete, or even completely missing. The main reason for this is that many such datasets are created by automatically harvesting, cleaning, and integrating data from heterogeneous (and sometimes noisy) primary sources. 

However, even when all the required metadata are available, in many cases the generation of the desired machine learning features involves time-consuming aggregation{\color{blue}s} and other processing tasks and may also be difficult to implement. For example, a number of data cleaning issues arise, for approaches using author-based features since 
author names have to be disambiguated in the case of
synonyms, or different spellings across publication venues.
Similarly, venue names might be recorded with different
forms (e.g., acronyms vs. full names). Such issues affect
the overall quality and, hence, the utility of these metadata.  

It is evident that, relying on rich article metadata is an important limitation for any machine learning approach to predict the expected impact of articles. On the other hand, an article's publication year is a basic information that is available in the vast majority of cases. As an indicative example, in the Crossref public data file of March 2020\footnote{\url{https://doi.org/10.13003/83B2GP}}, only $7.85\%$ of the records were missing this information. Moreover, due to the Initiative for Open Citations\footnote{\url{https://i4oc.org/}} (I4OC), an increasing number of publishers (with Elsevier being the most recent one) are committed to openly provide the reference lists of their articles. As a result, the majority of citation data are now available in open scholarly datasets (e.g., in Crossref). To summarize, the citations and the publication years of scientific articles are readily available
data.

Based on the above, we propose a set of features that can be easily calculated using article citations and publication years. In particular, we calculate the following:

\begin{itemize}
    \item \emph{cc\_total}: The total number of citations ever received by the article (i.e., its `citation count').
    \item \emph{cc\_1y}: The number of citations received by the article in the last year.
    \item \emph{cc\_3y}: The number of citations received by the article in the last $3$ years.
    \item \emph{cc\_5y}: The number of citations received by the article in the last $5$ years. 
\end{itemize}

The intuition behind these features is based on the idea of preferential attachment~\cite{barabasi2016network} and of its time-restricted version used in recent impact-based article ranking approaches~\cite{kanellos2020ranking}: articles that are likely to be highly cited
in the following few years are most likely those, which were
intensively cited in the recent past.

It should be noted that, although the minimum value of the features is zero in all cases, the largest value of each of them could be very diverse. This is why it is a good practice to normalize them before using them as input to the classifier.

\begin{table}[!t]
    \centering
    \small
    \begin{tabular}{ccc}
    \toprule
    \textbf{Sample set} & \textbf{Samples} & \textbf{Impactful samples} \\
    \midrule
    PMC $2011-2013$ ($3$ years)  & $229,207$ & $57,016 (24.88\%)$\\
    PMC $2011-2015$ ($5$ years)  & $229,207$  & $61,898 (27.01\%)$\\
    DBLP $2011-2013$ ($3$ years)  & $1,695,533$  & $387,506 (22.85\%)$ \\
    DBLP $2011-2015$ ($5$ years)  & $1,695,533$  & $339,351 (20.01\%)$ \\
    \bottomrule
    \end{tabular}
    \caption{Used sample sets}
    \label{tbl:samples}
\end{table}

\begin{table}[!t]
    \vspace{-0.5cm}
    \centering
    \small
    \begin{tabular}{lp{6cm}}
    \toprule
    \textbf{Classifier} & \textbf{Examined parameter values} \\
    \midrule
    LR \& cLR  &
    `max\_iter': $60, 80, 100, 120, 140, 160, 180, 200, 220, 240$ \newline
    `solver': `newton-cg', `lbfgs', `liblinear', `sag', `saga'\\
    \midrule
    DT \& cDT &
    `max\_depth': $1 - 32$ \newline
    `min\_samples\_split': $2, 5, 10, 20, 50, 100, 200$ \newline
    `min\_samples\_leaf': $1, 4, 7, 10$ \\
    \midrule
    RF \& cRF  &
    `max\_depth': $1,5,10,50$ \newline
    `n\_estimators': $100, 150, 200, 250, 300$ \newline
    `criterion': `gini', `entropy' \newline
    `max\_features': `log2', `sqrt'\\
    \bottomrule
    \end{tabular}
    \caption{Parameter values examined per classifier.}
    \vspace{-0.5cm}
    \label{tbl:grid}
\end{table}

\section{Evaluation}
\label{sec:eval}

\subsection{Setup}
\label{sec:eval-setup}

\noindent \textbf{Datasets.} For our experiments, we collected citations and publication years for scientific articles from two sources:
\begin{itemize}
    \item \emph{PMC}: The data were gathered from NCBI's PMC FTP directory\footnote{\url{ftp://ftp.ncbi.nlm.nih.gov/pub/pmc}} and are relevant to $1.12$ million open access scientific articles from life  sciences published between $1896$ and $2016$. Moreover, we removed data from the last year (they were incomplete, not the entire year was represented).
    \item \emph{DBLP}: The data were collected from AMiner's DBLP Citation Network dataset\footnote{\url{https://aminer.org/citation}}~\cite{Tang:08KDD} and are relevant to $3$ million scientific articles recorded by DBLP and published between $1936$ and $2018$. Moreover, we removed data from the last two, incomplete years.  
\end{itemize}
To create the labeled samples required for our analysis, we follow the hold-out evaluation approach~\cite{kanellos2019impact}: For each dataset we select the year $t=2010$ as a (virtual) present year and we split the dataset in two parts: the first one (articles published until $2010$, with $2010$ included) to calculate the feature vectors described in Section~\ref{sec:approach-feat} for all included articles; the second one to calculate the label for each sample, based on its future citations (see Section~\ref{sec:approach-prob}). We set $y=3$ and $y=5$ for the article impact future period (see
Section~\ref{sec:approach-prel}), which corresponds in 
both our datasets to the periods $2011-2013$, and
$2011-2015$, respectively. 
Table~\ref{tbl:samples} summarizes the statistics of the sample sets that have been created based on the aforementioned process.

\noindent \textbf{Classifiers.} We selected to use a set of well-known classifiers, along with their cost-sensitive versions\footnote{We used Scikit-learn's `balanced' mode for $class\_weight$ to automatically adjust weights inversely proportional to class frequencies in the input data.}. The reason we selected to include cost-sensitive versions is because they target the problem of imbalanced learning by using different misclassification costs for samples of different classes~\cite{he2013imbalanced}. 
As a result, we have configured and evaluated the following classification methods:

\begin{figure}[!t]
\includegraphics[scale=0.22]{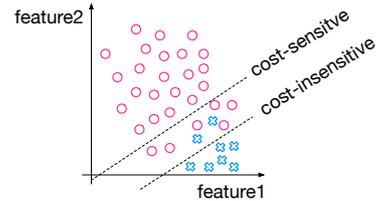}
\caption{Toy example showcasing why cost-sensitive approaches may achieve worse precision.}
\label{fig:example}
\end{figure}

\begin{itemize}
    \item \emph{LR}: Logistic regression
    \item \emph{cLR}: Cost-sensitive logistic regression
    \item \emph{DT}: Decision trees
    \item \emph{cDT}: Cost-sensitive decision trees
    \item \emph{RF}: Random forest
    \item \emph{cRF}: Cost-sensitive random forest
\end{itemize}

For all methods we used their Scikit-learn~\cite{scikit-learn} implementations and we have followed a two-fold, exhaustive grid search approach to identify the optimal values of their parameters according to the precision, recall, and F1 of the minority class. Table~\ref{tbl:grid} summarizes the parameter space examined, while Tables~\ref{tbl:pmc_conf}~\&~\ref{tbl:dblp_conf} in the Appendix enlist all the identified optimal configurations. Each optimal configuration is named as $[classifier]_{[measure]}$, where $[classifier]$ is the name of the corresponding classifier (e.g., LR, cLR) and $[measure]$ stands for the evaluation measure for which the configuration is optimal (e.g., `prec' for precision). 

\begin{table*}[h]
    \begin{subtable}[h]{0.45\textwidth}
        \centering
        \small
        \begin{tabular}{lcccc}
        \toprule
         &  \textbf{\small Precision} & \textbf{\small Recall} & \textbf{\small F1}\\
        \textbf{\small Classifier} & \textbf{\footnotesize(impactful|rest)} & \textbf{\footnotesize(impactful|rest)} & \textbf{\footnotesize(impactful|rest)}\\   
        \midrule
        LR$_{prec}$ & $\mathbf{0.85}|0.79$ & $0.23|0.99$ & $0.36|0.88$ \\
        LR$_{rec}$ & $\mathbf{0.85}|0.79$ & $0.23|0.99$ & $0.36|0.88$ \\
        LR$_{f1}$ & $\mathbf{0.85}|0.79$ & $0.23|0.99$ & $0.36|0.88$\\
        cLR$_{prec}$ & $0.57|0.85$ & $0.52|0.87$ & $0.55|0.86$ \\
        cLR$_{rec}$ & $0.57|0.85$ & $0.52|0.87$ & $0.55|0.86$  \\
        cLR$_{f1}$ & $0.57|0.85$ & $0.52|0.87$ & $0.55|0.86$  \\       
        DT$_{prec}$ & $0.66|0.82$ & $0.38|0.93$ & $0.48|0.87$ \\
        DT$_{rec}$ & $0.66|0.82$ & $0.38|0.93$ & $0.48|0.87$\\
        DT$_{f1}$ & $0.66|0.82$ & $0.38|0.93$ & $0.48|0.87$\\
        cDT$_{prec}$ & $0.60|0.85$ & $0.52|0.89$ & $\mathbf{0.56}|0.87$ \\
        cDT$_{rec}$ & $0.50|0.87$ & $0.63|0.79$ & $\mathbf{0.56}|0.83$ \\
        cDT$_{f1}$ & $0.52|0.86$ & $0.60|0.81$ & $0.55|0.84$ \\
        RF$_{prec}$ & $0.70|0.82$ & $0.38|0.95$ & $0.50|0.88$ \\
        RF$_{rec}$ & $0.71|0.82$ & $0.37|0.95$ & $0.48|0.88$ \\
        RF$_{f1}$ & $0.71|0.82$ & $0.36|0.95$ & $0.48|0.88$ \\
        cRF$_{prec}$ & $0.56|0.85$ & $0.53|0.86$ & $0.54|0.85$ \\
        cRF$_{rec}$ & $0.47|0.87$ & $\mathbf{0.65}|0.76$ & $0.55|0.81$\\
        cRF$_{f1}$ & $0.48|0.87$ & $\mathbf{0.65}|0.77$ & $0.55|0.81$\\
        \bottomrule
        \end{tabular}
        \caption{PMC}
        \label{tbl:pmc_res_y3}
    \end{subtable}
    \hfill
    \begin{subtable}[h]{0.45\textwidth}
       \centering
       \small
        \begin{tabular}{lcccc}
        \toprule
         &  \textbf{\small Precision} & \textbf{\small Recall} & \textbf{\small F1}\\
        \textbf{\small Classifier} & \textbf{\footnotesize(impactful|rest)} & \textbf{\footnotesize(impactful|rest)} & \textbf{\footnotesize(impactful|rest)}\\   
        \midrule
        LR$_{prec}$ & $\mathbf{0.97}|0.82$ & $0.25|1.00$ & $0.39|0.90$ \\
        LR$_{rec}$ & $0.96|0.82$ & $0.26|1.00$ & $0.40|0.90$ \\
        LR$_{f1}$ & $0.96|0.82$ & $0.25|1.00$ & $0.40|0.90$ \\
        cLR$_{prec}$ & $0.70|0.88$ & $0.57|0.93$ & $0.63|0.90$ \\
        cLR$_{rec}$ & $0.70|0.88$ & $0.57|0.93$ & $0.63|0.90$ \\
        cLR$_{f1}$ & $0.71|0.88$ & $0.56|0.93$ & $0.63|0.90$ \\       
        DT$_{prec}$ & $0.80|0.88$ & $0.55|0.96$ & $\mathbf{0.65}|0.92$ \\
        DT$_{rec}$ & $0.72|0.89$ & $0.61|0.93$ & $0.61|0.91$ \\
        DT$_{f1}$ & $0.72|0.89$ & $0.61|0.93$ & $0.61|0.91$ \\
        cDT$_{prec}$ & $0.58|0.92$ & $0.74|0.84$ & $\mathbf{0.65}|0.88$ \\
        cDT$_{rec}$ & $0.52|0.93$ & $\mathbf{0.79}|0.78$ & $0.63|0.85$ \\
        cDT$_{f1}$ & $0.58|0.92$ & $0.75|0.84$ & $\mathbf{0.65}|0.88$\\
        RF$_{prec}$ & $0.72|0.88$ & $0.56|0.94$ & $0.63|0.91$ \\
        RF$_{rec}$ & $0.72|0.88$ & $0.56|0.94$ & $0.63|0.91$ \\
        RF$_{f1}$ & $0.77|0.87$ & $0.54|0.95$ & $0.63|0.91$\\
        cRF$_{prec}$ & $0.64|0.89$ & $0.63|0.89$ & $0.64|0.89$ \\
        cRF$_{rec}$ & $0.57|0.92$ & $0.76|0.83$ & $\mathbf{0.65}|0.87$\\
        cRF$_{f1}$ & $0.58|0.92$ & $0.76|0.84$ & $0.65|0.88$ \\
        \bottomrule
        \end{tabular}
        \caption{DBLP}
        \label{tbl:dblp_res_y3}
     \end{subtable}
     \caption{Precision, recall, and F1 based on future citations in [2011-2013] (3 years). Tables~\ref{tbl:pmc_conf}~\&~\ref{tbl:dblp_conf} contain the detailed configuration for each examined classifier.}
     \label{tbl:res_y3}
\end{table*}

\begin{table*}[h]
    \begin{subtable}[h]{0.45\textwidth}
        \centering
        \small
        \begin{tabular}{lcccc}
        \toprule
         &  \textbf{\small Precision} & \textbf{\small Recall} & \textbf{\small F1}\\
        \textbf{\small Classifier} & \textbf{\footnotesize(impactful|rest)} & \textbf{\footnotesize(impactful|rest)} & \textbf{\footnotesize(impactful|rest)}\\   
        \midrule
        LR$_{prec}$ & $\mathbf{0.89}|0.78$ & $0.26|0.99$ & $0.40|0.87$ \\
        LR$_{rec}$ & $\mathbf{0.89}|0.78$ & $0.26|0.99$ & $0.40|0.87$ \\
        LR$_{f1}$ & $\mathbf{0.89}|0.78$ & $0.25|0.99$ & $0.39|0.87$\\
        cLR$_{prec}$ & $0.60|0.82$ & $0.49|0.88$ & $0.54|0.85$ \\
        cLR$_{rec}$ & $0.60|0.82$ & $0.48|0.88$ & $0.54|0.85$  \\
        cLR$_{f1}$ & $0.60|0.82$ & $0.49|0.88$ & $0.54|0.85$  \\       
        DT$_{prec}$ & $0.75|0.81$ & $0.38|0.95$ & $0.50|0.87$ \\
        DT$_{rec}$ & $0.75|0.80$ & $0.35|0.96$ & $0.48|0.87$\\
        DT$_{f1}$ & $0.75|0.81$ & $0.39|0.95$ & $0.51|0.87$\\
        cDT$_{prec}$ & $0.60|0.82$ & $0.49|0.88$ & $0.54|0.85$ \\
        cDT$_{rec}$ & $0.50|0.84$ & $0.61|0.78$ & $0.55|0.81$ \\
        cDT$_{f1}$ & $0.53|0.84$ & $0.60|0.81$ & $\mathbf{0.56}|0.82$ \\
        RF$_{prec}$ & $0.72|0.80$ & $0.37|0.95$ & $0.49|0.87$ \\
        RF$_{rec}$ & $0.73|0.81$ & $0.41|0.95$ & $0.53|0.87$ \\
        RF$_{f1}$ & $0.74|0.81$ & $0.41|0.95$ & $0.52|0.87$ \\
        cRF$_{prec}$ & $0.57|0.82$ & $0.49|0.86$ & $0.52|0.84$ \\
        cRF$_{rec}$ & $0.50|0.84$ & $\mathbf{0.61}|0.77$ & $0.55|0.81$\\
        cRF$_{f1}$ & $0.50|0.84$ & $\mathbf{0.61}|0.77$ & $0.55|0.81$\\
        \bottomrule
        \end{tabular}
        \caption{PMC}
        \label{tbl:pmc_res_y5}
    \end{subtable}
    \hfill
    \begin{subtable}[h]{0.45\textwidth}
       \centering
       \small
        \begin{tabular}{lcccc}
        \toprule
         &  \textbf{\small Precision} & \textbf{\small Recall} & \textbf{\small F1}\\
        \textbf{\small Classifier} & \textbf{\footnotesize(impactful|rest)} & \textbf{\footnotesize(impactful|rest)} & \textbf{\footnotesize(impactful|rest)}\\   
        \midrule
        LR$_{prec}$ & $0.96|0.84$ & $0.24|1.00$ & $0.39|0.91$ \\
        LR$_{rec}$ & $0.96|0.84$ & $0.24|1.00$ & $0.39|0.91$ \\
        LR$_{f1}$ & $\mathbf{0.97}|0.84$ & $0.24|1.00$ & $0.38|0.91$ \\
        cLR$_{prec}$ & $0.70|0.90$ & $0.61|0.93$ & $\mathbf{0.65}|0.92$ \\
        cLR$_{rec}$ & $0.73|0.90$ & $0.58|0.94$ & $\mathbf{0.65}|0.92$ \\
        cLR$_{f1}$ & $0.70|0.90$ & $0.60|0.93$ & $\mathbf{0.65}|0.92$ \\       
        DT$_{prec}$ & $0.87|0.87$ & $0.42|0.98$ & $0.56|0.92$ \\
        DT$_{rec}$ & $0.73|0.90$ & $0.56|0.95$ & $0.63|0.92$ \\
        DT$_{f1}$ & $0.77|0.89$ & $0.52|0.96$ & $0.62|0.92$ \\
        cDT$_{prec}$ & $0.59|0.93$ & $0.72|0.88$ & $\mathbf{0.65}|0.90$ \\
        cDT$_{rec}$ & $0.47|0.94$ & $\mathbf{0.82}|0.77$ & $0.60|0.85$ \\
        cDT$_{f1}$ & $0.59|0.93$ & $0.72|0.88$ & $\mathbf{0.65}|0.90$\\
        RF$_{prec}$ & $0.83|0.89$ & $0.52|0.97$ & $0.64|0.93$ \\
        RF$_{rec}$ & $0.74|0.90$ & $0.56|0.95$ & $0.64|0.92$ \\
        RF$_{f1}$ & $0.80|0.90$ & $0.56|0.96$ & $0.66|0.93$\\
        cRF$_{prec}$ & $0.62|0.91$ & $0.66|0.90$ & $0.64|0.91$ \\
        cRF$_{rec}$ & $0.59|0.91$ & $0.67|0.89$ & $0.63|0.90$\\
        cRF$_{f1}$ & $0.55|0.93$ & $0.76|0.84$  & $0.64|0.89$\\
        \bottomrule
        \end{tabular}
        \caption{DBLP}
        \label{tbl:dblp_res_y5}
     \end{subtable}
     \caption{Precision, recall, and F1 based on future citations in [2011-2015] (5 years). Tables~\ref{tbl:pmc_conf}~\&~\ref{tbl:dblp_conf} contain the detailed configuration for each examined classifier.}
     \label{tbl:res_y5}
\end{table*}

\subsection{Results}
\label{sec:eval-result}

Because of the imbalanced nature of the classification problem we study, it is very important to carefully select the measures that will be used for the evaluation of the effectiveness of the examined approaches. For example, as it was discussed in Section~\ref{sec:approach-prob}, \emph{accuracy} that is commonly used for generic classification approaches, is not a good option, since it is mostly affected by the misclassification of samples from the majority class. However, in most imbalanced problems, like the one we have here, the minority class has the most importance.

Therefore, we do not report the accuracy of the examined approaches. In any case, all configurations achieved accuracy between $0.73$ and $0.99$. Following the best practices for the evaluation of imbalanced classification approaches, we instead measure the precision, recall, and F1 of the minority class. We indicatively report the same measures for the majority class, as well. However our main objective is to perform well according to the measures calculated for the minority class. Note that, each of these three measures may be preferable for different applications.

Tables~\ref{tbl:dblp_res_y3}~\&~\ref{tbl:dblp_res_y5} summarize the results of the performed experiments. The results are very similar for both data sets (PMC and DBLP) and for both values of the parameter $y$. A general observation is that, when we focus on precision, cost-insensitive classification approaches 
perform adequately well and, thus, there is no need to work with cost-sensitive versions. However, the same experiments highlight that the latter approaches can significantly improve the effectiveness based on the recall and F1. 

This behavior is not surprising: By default, in several classifiers, the optimization process targets at accuracy maximization, since all samples equally contribute to the loss function to be minimized. Consequently, in areas of the hyperspace where the samples of different classes are not easily separable, the samples of the majority class are favored (i.e., correctly classified) due to their dominance in numbers. Consider, for instance, the two minority class samples (cross marks) and the six majority class ones (cyclic marks) between the two alternative hyperplanes of the toy example in Figure~\ref{fig:example}: Classifying all of them to the majority class would induce $3$~times less cost to the classifier than classifying them to the minority class. In this way the cost-insensitive classifier also achieves good precision for the minority class (no false positives in this example). The drawback is that this results in many false negatives for the minority class (the most important one). Cost-sensitive approaches alleviate this issue improving the recall and F1 of the minority class, with the counter-effect of a larger number of false positives for the minority class.  

Focusing on the differences of the examined classification approaches, it seems that cost-insensitive Logistic Regression is, by far, the best option for applications focusing on precision, achieving values between $0.85$ and $0.97$ for all datasets. However, this is achieved by allowing very significant losses in recall and F1 (values below $0.27$ and $0.41$ for all datasets, respectively). On the other hand, cost-sensitive Random Forest and Decision Tree classifiers seem to be the best options when recall and F1 are more important (albeit their losses in precision are significant).

\section{Related work}
\label{sec:related}

The vast majority of works that attempt to estimate the expected impact of scientific articles focus on predicting the exact number of citations each article will receive in a given future period, a problem know as \emph{Citation Count Prediction} (CCP). Most of these works incorporate a wide range of features based on the article's content, novelty, author list, venue, topic, citations, reviews, to name only a few. The corresponding predicting models are based on various regression models like Linear Regression~\cite{yan2011citation,Zhu2018CitationCP}, k-NN~\cite{yan2011citation}, SVR~\cite{yan2011citation,livne2013predicting,Zhu2018CitationCP,Li2015TrendBasedCC}, Gaussian Process Regression~\cite{yan2012better}, CART Model~\cite{yan2012better,yan2011citation}, ZINB Regression~\cite{didegah2013determinants}, or various types of neural networks~\cite{Liu2020CitationCP,Zhu2018CitationCP,Wen2020PaperCC,abrishami2019predicting,Li2019ADL,Li2019ANC}. In most works, one or more regression models are tested on the complete data set, with the notable exception of~\cite{Li2015TrendBasedCC}, which first attempts to identify the current citation trend of each article (e.g., early burst, no burst, late burst, etc) and then applies a different model for each case. As it was elaborated in Section~\ref{sec:approach-prob}, CCP is a very difficult problem and there are many, not easily quantified factors that can significantly affect the performance of such approaches. Also, such approaches rely on article metadata that are difficult to collect and that they should be undergo complex to implement and time-consuming processing (see also Section~\ref{sec:approach-feat}).   

In another line of work, based on the fact that co-authorship and citation-based features seemed to be effective for earlier approaches, the authors of \cite{pobiedina2016citation} follow a link-prediction-inspired approach to solve CCP. They also investigate the effectiveness of their approach in a relevant classification problem based on a set of arbitrarily determined classes. However, training their approach requires a heavy pattern mining analysis of the underlying citation network and also considers author- and venue-based features, which face the already discussed issues. It should be noted that there are also some link prediction approaches that aim to reveal missing citations between a set of articles (e.g.,\cite{Yu2012CitationPI}), these approaches are irrelevant to the problem of impact prediction though. Furthermore, in~\cite{su2020prediction} an impact-based classification problem is studied, but the features of the proposed approach rely on difficult to collect article metadata (e.g., information about academic and funding organizations). As a result, this approach cannot be easily used in practice. 
Finally, there are methods that attempt to estimate the rank of articles based on their expected impact. A thorough survey and experimental study of such methods can be found in~\cite{kanellos2019impact}. This problem is easier than CCP, since only the partial ordering of the articles according to their expected impact should be estimated, but it is still more difficult than the problem we focus on.

\section{Conclusion}

In this work, we propose a simplified approach that can significantly simplify the work of researchers and developers working on applications that rely on the prediction of the expected impact of scientific articles. The proposed approach is based on classifying the articles in two categories (‘impactful’ / ‘impactless’) based on a set of features that can be calculated using a minimal set of article metadata. Furthermore, we experimentally evaluated this approach using various well-established classifiers showing that the results are more than adequate. The aforementioned experiments have been performed with caution taking into account the imbalanced nature of the classification problem at hand. 

In the future, we plan to further investigate the imbalanced nature of the problem by examining other approaches like methods that perform over-sampling of the minority class, others that perform under-sampling of the majority class, or methods combining these two approaches (e.g., SMOTEEN). Additionally, we plan to examine a wider range of parameters for the examined approaches, for instance, examining a range of custom weights for cost-sensitive approaches. Finally, we plan to take full advantage of the Head/Tail Breaks approach to study a non-binary version of the classification problem.


\bibliographystyle{ACM-Reference-Format}
\bibliography{sample-base}

\appendix

\section{Used parameter configurations}

Tables~\ref{tbl:pmc_conf}~\&~\ref{tbl:dblp_conf} summarize the configuration for each used approach. The names of the parameters are based on the input parameters of the corresponding Scikit-learn functions. Omitted input parameters were not configured (their default values had been selected). 

\begin{table*}[h]
        \centering
        \small
        \begin{tabular}{l p{7.2cm} p{7.2cm} }
        \toprule
        \textbf{Classifier} & \textbf{Configuration for $y=3$} & \textbf{Configuration for $y=5$}\\   
        \midrule
        LR$_{prec}$ & `max\_iter': 200, `solver': `sag' & `max\_iter': 160, `solver': `sag'\\
        LR$_{rec}$ & `max\_iter': 80, `solver': `sag' & `max\_iter': 80, `solver': `sag' \\
        LR$_{f1}$ & `max\_iter': 180, `solver': `sag' & `max\_iter': 240, `solver': `sag'\\
        cLR$_{prec}$ & `max\_iter': 100, `solver': `sag' & `max\_iter': 60, `solver': `sag'\\
        cLR$_{rec}$ & `max\_iter': 120, `solver': `sag' & `max\_iter': 140, `solver': `sag' \\
        cLR$_{f1}$ & `max\_iter': 180, `solver': `sag' & `max\_iter': 140, `solver': `sag'\\            
        DT$_{prec}$ & `max\_depth': 3, `min\_samples\_leaf': 1,  `min\_samples\_split': 2 & `max\_depth': 4, `min\_samples\_leaf': 1, `min\_samples\_split': 2\\
        DT$_{rec}$ & `max\_depth': 1, `min\_samples\_leaf': 1, `min\_samples\_split': 2 & `max\_depth': 3, `min\_samples\_leaf': 1, `min\_samples\_split': 2 \\
        DT$_{f1}$ & `max\_depth': 1, `min\_samples\_leaf': 1, `min\_samples\_split': 2 & `max\_depth': 8, `min\_samples\_leaf': 10, `min\_samples\_split': 200\\
        cDT$_{prec}$ & `max\_depth': 1, `min\_samples\_leaf': 1, `min\_samples\_split': 2 & `max\_depth': 1, `min\_samples\_leaf': 1, `min\_samples\_split': 2 \\
        cDT$_{rec}$ & `max\_depth': 2, `min\_samples\_leaf': 1, `min\_samples\_split': 2 & `max\_depth': 2, `min\_samples\_leaf': 1, `min\_samples\_split': 2 \\
        cDT$_{f1}$ & `max\_depth': 7, `min\_samples\_leaf': 4, `min\_samples\_split': 20 & `max\_depth': 7, `min\_samples\_leaf': 4, `min\_samples\_split': 50 \\
        RF$_{prec}$ & `criterion': `gini', `max\_depth': 1, `max\_features': `log2', \newline `n\_estimators': 200 & `criterion': `gini', `max\_depth': 1, `max\_features': `log2', \newline `n\_estimators': 200 \\
        RF$_{rec}$ & `criterion': `gini', `max\_depth': 10, `max\_features': `log2', \newline `n\_estimators': 300 & `criterion': `gini', `max\_depth': 10, `max\_features': `sqrt', \newline `n\_estimators': 300 \\
        RF$_{f1}$ & `criterion': `entropy', `max\_depth': 10, `max\_features': `sqrt', \newline `n\_estimators': 200 & `criterion': `entropy', `max\_depth': 10, `max\_features': `sqrt', \newline 'n\_estimators': 300 \\
        cRF$_{prec}$ & `criterion': `entropy', `max\_depth': 1, `max\_features': `log2', \newline `n\_estimators': 150 & `criterion': 'entropy', `max\_depth': 1, `max\_features': 'log2', \newline `n\_estimators': 100 \\
        cRF$_{rec}$ & `criterion': `gini', `max\_depth': 5, `max\_features': `sqrt', \newline `n\_estimators': 150 & `criterion': `entropy', `max\_depth': 5, `max\_features': `log2', \newline `n\_estimators': 100 \\
        cRF$_{f1}$ & `criterion': `entropy', `max\_depth': 10, `max\_features': `log2', \newline `n\_estimators': 150 & `criterion': `gini', `max\_depth': 5, `max\_features': `sqrt', `n\_estimators': 300 \\
        \bottomrule
        \end{tabular}
        \caption{Parameter configurations for PMC.}
        \label{tbl:pmc_conf}
\end{table*}

\begin{table*}[h]
        \centering
        \small
        \begin{tabular}{l p{7.2cm} p{7.2cm} }
        \toprule
        \textbf{Classifier} & \textbf{Configuration for $y=3$} & \textbf{Configuration for $y=5$}\\   
        \midrule
        LR$_{prec}$ & `max\_iter': 80, `solver': `sag' & `max\_iter': 100, `solver': 'sag' \\
        LR$_{rec}$ &  `max\_iter': 80, `solver': `sag' & `max\_iter': 140, `solver': 'sag' \\
        LR$_{f1}$ & `max\_iter': 220, `solver': `saga'& `max\_iter': 220, `solver': `sag' \\
        cLR$_{prec}$ & `max\_iter': 200, `solver': `sag'& `max\_iter': 180, `solver': `sag' \\
        cLR$_{rec}$ & `max\_iter': 140, `solver': `sag'& `max\_iter': 160, `solver': `sag' \\
        cLR$_{f1}$ & `max\_iter': 100, `solver': `sag' & `max\_iter': 60, `solver': `newton-cg' \\       
        DT$_{prec}$ & `max\_depth': 6, `min\_samples\_leaf': 1, `min\_samples\_split': 2 & `max\_depth': 3, `min\_samples\_leaf': 1, `min\_samples\_split': 2 \\
        DT$_{rec}$ & `max\_depth': 3, `min\_samples\_leaf': 1, `min\_samples\_split': 2 & `max\_depth': 1, `min\_samples\_leaf': 1, `min\_samples\_split': 2 \\
        DT$_{f1}$ &  `max\_depth': 3, `min\_samples\_leaf': 1, `min\_samples\_split': 2 & `max\_depth': 4, `min\_samples\_leaf': 1, `min\_samples\_split': 2 \\
        cDT$_{prec}$ & `max\_depth': 14, `min\_samples\_leaf': 10, `min\_samples\_split': 2 & `max\_depth': 4, `min\_samples\_leaf': 1, `min\_samples\_split': 2 \\
        cDT$_{rec}$ & `max\_depth': 2, `min\_samples\_leaf': 1, `min\_samples\_split': 2 & `max\_depth': 2, `min\_samples\_leaf': 1, `min\_samples\_split': 2 \\
        cDT$_{f1}$ &  `max\_depth': 11, `min\_samples\_leaf': 10, `min\_samples\_split': 200 & `max\_depth': 4, `min\_samples\_leaf': 1, `min\_samples\_split': 2 \\
        RF$_{prec}$ & `criterion': `entropy', `max\_depth': 1, `max\_features': `log2', \newline `n\_estimators': 150 & `criterion': `gini', `max\_depth': 5, `max\_features': `sqrt', `n\_estimators': 100\\
        RF$_{rec}$ & `criterion': `entropy', `max\_depth': 1, `max\_features': `log2', \newline `n\_estimators': 150 & `criterion': `entropy', `max\_depth': 1, `max\_features': `log2', `n\_estimators': 150 \\
        RF$_{f1}$ & `criterion': `gini', `max\_depth': 5, `max\_features': `log2', \newline `n\_estimators': 100 & `criterion': `entropy', `max\_depth': 10, `max\_features': `sqrt', `n\_estimators': 250 \\
        cRF$_{prec}$ & `criterion': `entropy', `max\_depth': 1, `max\_features': `log2', `n\_estimators': 250 & `criterion': `entropy', `max\_depth': 1, `max\_features': `log2', `n\_estimators': 100 \\
        cRF$_{rec}$ & `criterion': `gini', `max\_depth': 5, `max\_features': `log2', `n\_estimators': 100 & `criterion': `gini', `max\_depth': 1, `max\_features': `log2', `n\_estimators': 150 \\
        cRF$_{f1}$ & `criterion': `entropy', `max\_depth': 10, `max\_features': `log2', `n\_estimators': 150 & `criterion': `entropy', `max\_depth': 10, `max\_features': `sqrt', `n\_estimators': 150 \\
        \bottomrule
        \end{tabular}
        \caption{Parameter configurations for DBLP.}
        \label{tbl:dblp_conf}
\end{table*}

\end{document}